# MEASUREMENTS OF THE BEAM LONGITUDINAL PROPERTIES IN THE FERMILAB LINAC*

R. Sharankova[†,1], J-P. Carneiro[1], E. Chen[1], A. Shemyakin[1]
[1]Fermi National Accelerator Laboratory, Batavia, IL, USA

## Abstract

The Fermilab Linac delivers 400MeV, 25mA H⁻ beam. The longitudinal bunch parameters are reconstructed using a Bunch Shape Monitor (BSM) installed in the middle of the Linac. For that, the bunch length is measured as a function of the phase of an upstream cavity and fitted to simulations. The cavity gradient and its phase with respect to the beam are recovered from readings of Beam Position Monitors (BPMs). Since the cavity provides a significant transverse defocusing, the BSM measurements are correlated with transverse beam size measurements by a wire scanner (WS). Simulations connect these three types of measurements, allowing to deduce the longitudinal emittance and Courant-Snyder parameters.

## INTRODUCTION

The Fermilab Linac consists of a drift-tube Linac (DTL) and a side-coupled Linac (SCL) with a Transition section between them. The DTL is made up of 5 Alvarez RF cavities with 201.49 MHz resonant frequency which accelerate beam to 116.5 MeV design energy. The SCL consists of 7 RF cavities, operates at 805 MHz resonant frequency and accelerates beam to 401.5 MeV design energy. The transition section has two bunching cavities and is responsible for matching the DTL and SCL sections in all planes. It is well instrumented with diagnostics and is well-suited for extracting emittance measurements in the longitudinal and transverse planes. Transverse parameter reconstruction is described in Ref. [1], [2], and this paper focuses on the longitudinal properties.

### The Linac Transition section

Figure 1 shows a schematic of the Transition section. The beamline components in the section include 3 wire scanners (WS, shown in black), 4 quadrupole magnets in a FODO configuration (dark green), 2 bunching RF cavities (the larger Buncher followed by the smaller Vernier, in blue), 1 bunch shape monitor (BSM, brown), and 3 beam position monitors (BPMs) (not shown explicitly). $D_{ij}$ in an element name indicate location as shown in Fig. 1.

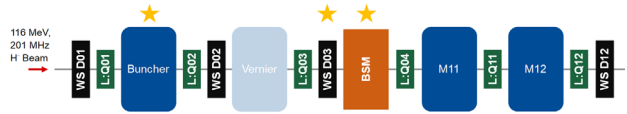

Figure 1: A schematic of the Transition section. Star indicates elements used in the measurements discussed in this work.

### Beam Position Monitors

The Transition section has 4 two-planes button BPMs located directly in front of every quadrupole magnet. Beside horizontal and vertical position of the beam centroid, the BPMs measure relative phase w.r.t. the RF reference line. The BPMs operate at the second beam harmonic of 402.5 MHz. Reported phase is not absolutely calibrated for cable and electronics delay, so it is only used in relative measurements.

### Wire Scanners

The Linac employs wire scanners with 3 wires (horizontal, vertical and 45 degrees) fixed to a frame which runs through beam at a constant speed to record beam profiles in the horizontal (x), vertical (y) and 45-degree (u) planes. More details on the wire scanners can be found in [1].

### Bunch Shape monitor

The bunch shape monitor (BSM) [3] is a device that measures the longitudinal profile of the bunch. Figure 2 shows a schematic of the BSM working principle.

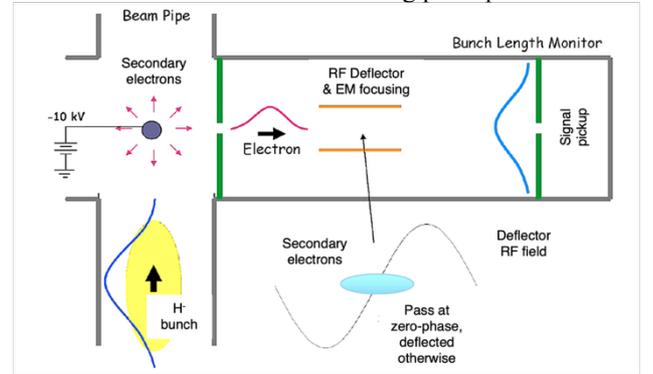

Figure 2: BSM working principle

A tungsten wire at negative electrical potential is moved in the H- beam path. The ions hitting the wire produce secondary electrons which are repelled by the wire potential towards the primary collimator (slit). The secondary electrons carry the time (longitudinal) structure of the H-bunch. The electrons enter an RF deflector cavity where only those at zero-phase w.r.t. the cavity pass undeflected towards the secondary slit. Finally, the electrons are collected by an electron multiplier tube and recorded as electrical signal. The signal magnitude is proportional to the electron density. Scanning the RF phase of the deflector cavity allows to measure the electrons density along the time (longitudinal) dimension of the bunch.

## MEASUREMENTS

### Cavity phasing

Cavity phase scans are performed to define the beam synchronous phase as well as RF cavity amplitude calibration in the traditional way. The procedure is as follows: all cavities downstream of the cavity of interest are turned off to create a drift region. The cavity phase is scanned 360 degrees while the beam phases $\{\varphi_i\}$ measured by BPMs downstream are recorded. Then, the difference of between phase of the i-th and 0-th BPM are calculated $\Delta\varphi_i = \varphi_i - \varphi_0$. By itself, this value doesn't bring useful information since there is no absolute calibration of the phases. However, dependence of $\{\Delta\varphi_i\}$ on the cavity phase is meaningful.

Let $\{\Delta\Phi_i\}$ be actual phase differences as they would in the case of absolute BPM phase calibration:

$$\Delta\Phi_i = 2\pi f_{BPM} \frac{L_i}{v} \quad (1)$$

where $L_i$ is the distance to the i-th BPM from BPM #0, $f_{BPM}$ is the BPM frequency (402.5 MHz), and $v$ is the beam velocity. For small deviations $\delta v$ of the velocity from its value $v_0$ when the cavity if off,

$$\frac{\delta v}{v_0} = -\frac{\delta\Phi_i}{\Delta\Phi_{0i}} \quad (2)$$

where $\Delta\Phi_{0i}$ are the phase differences at $v = v_0$, and $\delta\Phi_i = \Delta\Phi_i - \Delta\Phi_{0i}$. Because of the double subtraction, $\{\delta\Phi_i\}$ do not depend on the absolute calibration and are equal to the corresponding measured differences $\{\delta\varphi_i\}$. Since the input energy $E_{0k}$ is known with a good enough accuracy, $\{\Delta\Phi_{0i}\}$ are calculated from Eq.(1). Then, the variation of kinetic energy $\delta E_k$ is

$$\frac{\delta E_k}{E_{0k}} = \gamma(\gamma + 1)\frac{\delta v}{v_0}, \quad \gamma = \sqrt{1-\beta^2}, \quad \beta = \frac{v_0}{c} \quad (3)$$

where $c$ is speed of light.

As simulations show, the energy gain in the relatively small cavities of the Transition section can be well approximated by a cosine function:

$$\delta E_k = V_0 \cos\varphi_{beam} \quad (4)$$

where $\varphi_{beam}$ is the phase of the beam relative to the phase of the maximum acceleration. Therefore, the recipe is to fit the measured BPM phase differences to a cosine function of the cavity phase $\varphi_{cavity}$, find the cavity phase offset $\varphi_m$ corresponding to maximum acceleration as the point where $\delta\varphi_i(\varphi_{beam})$ reaches minimum, and determine the cavity amplitude $V_0$ from Eq.(1)- Eq.(4). After that, $\varphi_{beam} = \varphi_{cavity} - \varphi_m$.

To phase the Buncher, a longitudinal drift space is created after it by turning off the Vernier. Figure 3 shows an example of the cosine fitting for the Buncher cavity for several BPMs. The value of $\varphi_m$ (285°) is consistent between the curves. The amplitude of curves in Fig. 3 change linearly with the distance (Fig. 4). The slope of the linear fit is proportional to the cavity amplitude:

$$\frac{d\varphi}{dz} = \frac{2\pi f_{BPM}}{c \cdot E_{0k}} \frac{V_0}{\gamma \cdot (\gamma + 1)}. \quad (5)$$

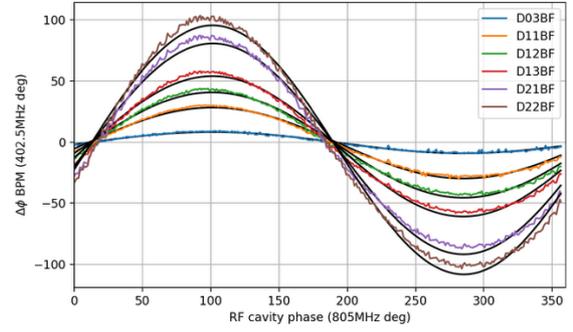

Figure 3: Buncher cavity phase scan (deg of 805MHz).

The slope of Fig. 4 corresponds to a cavity amplitude of 4.7 MeV. This value is implemented in following TraceWin [4] simulations.

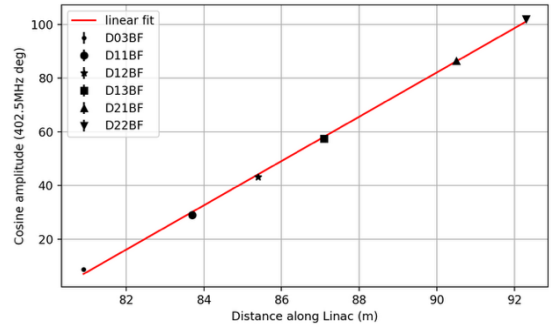

Figure 4: Amplitudes of cosine fits in Fig.3 vs BPM positions. Linear fit is applied to extract slope parameter.

### Reconstruction of longitudinal parameters

The same Buncher cavity phase scan with Vernier off is repeated while recording the longitudinal profiles with BSM. Figure 5 shows examples of such measurements.

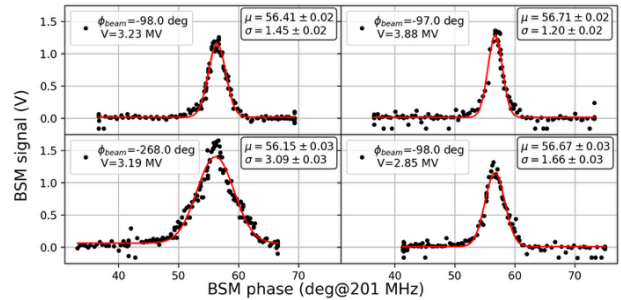

Figure 5: Longitudinal bunch profiles at different Buncher cavity phase and gradient settings.

The bunch length is extracted by fitting a Gaussian function. It is minimal close to $\varphi_{beam} = -90$ deg per Buncher cavity design.

The data were then fitted using TraceWin [4] using the Buncher parameters determined in measurements with BPMs (Fig.6). Resulting Twiss parameters and emittance a are reported in [2] and in Fig. 6. The rms normalized emittance is 1 mm·mrad, and Twiss functions are $\beta_z = 8\ m$, $\alpha_z = -0.5$.

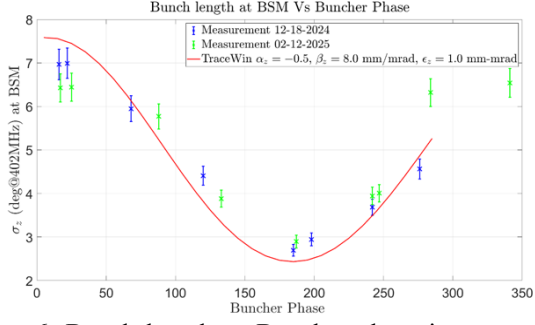

Figure 6: Bunch length vs Buncher phase in two sets of data compared with simulation.

The described procedure is analogous to quadrupole scans. However, Fig. 6 might be misleading since the focusing strength is proportional to $V_0 \sin \varphi_{beam}$, and portions of the plot symmetrical with respect to $\varphi_m$ represent the same focusing. In Fig. 7, the data are plotted against $V_0 \sin \varphi_{beam}$. It should be pointed out that $V_0$ here is in MV and not in MeV. The bunch size doesn't reach a minimum, which affect accuracy of reconstructing the beam parameters. Note that the points in the left portion of Fig. 7 were added by increasing the Buncher field gradient at $\varphi_{beam} \sim -90°$. The RF regulation for the Buncher could not sustain higher gradients which is we were unable to scan the beam size to the minimum and overfocusing, as is typically done in quadrupole scans.

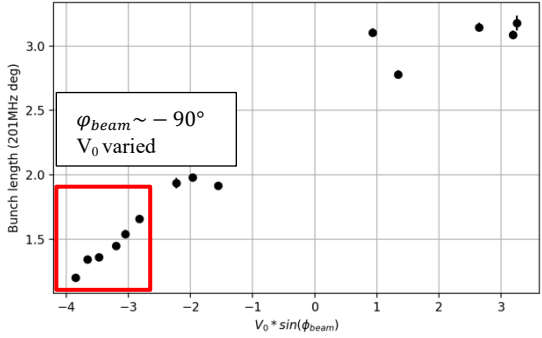

Figure 7: Bunch length vs. RF focusing strength.

*Transverse RF de-focusing*

Cavity RF fields affect both longitudinal and transverse motion of the ions. Coordinates for description of an off-center particle motion can be chosen similar in all planes, (x, x'), (y, y'), (z, z'), where apostrophe means derivation along the trajectory. With this choice, when a cavity can be described as a thin axisymmetric lens in all planes, it can be characterized by the focal lengths $f_{x,y,z}$:

$$\frac{1}{f_z} = \frac{2\pi f_c}{c} \cdot \frac{V_0}{Mc^2} \cdot \frac{1}{\beta^3 \gamma^3} \sin \varphi_{beam} = -\frac{2}{f_{x,y}};$$

$$\frac{1}{f_x} \equiv \frac{\Delta x'}{x}, \frac{1}{f_y} \equiv \frac{\Delta y'}{y}, \frac{1}{f_z} \equiv \frac{\Delta z'}{z}, \quad (5)$$

where $f_c$ is the cavity frequency, and $M$ is the ion mass. Eq. (5) indicates that transverse and longitudinal focusing have opposite sign, i.e. the cavity phase focusing the beam longitudinally, it defocuses transversely. While cavities are not optically thin, one can expect the connection between the planes following Eq. (5).

To measure this effect, the transverse profiles are recorded using the wire scanner WS D03, located just upstream of the BSM. The profiles are then fit with a Gaussian function to extract the bunch size in x and y.

Fig. 8 compares the changes in the bunch sizes in all planes vs the Buncher cavity phase. Effect of transverse defocusing is significant. Smallest transverse sizes are reached at the cavity phases corresponding the longest bunch length.

Note that the transverse de-focusing effect is clearly visible in the longitudinal measurements as a change in the integrals over the BSM curves as in Fig. 5.

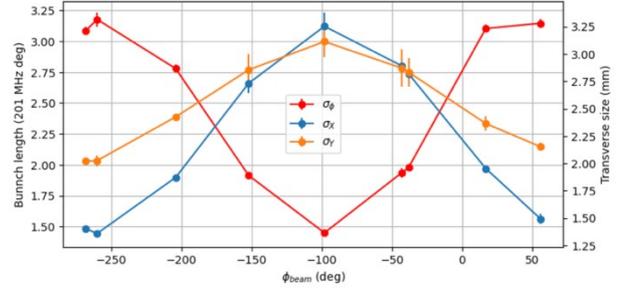

Figure 8: Longitudinal and transverse bunch size vs Buncher phase (shown with respect to the beam).

Fig. 9 shows comparison of the measured beam size in the vertical (y) plane with TraceWin simulations. The simulations correctly capture the trend in the measurements but disagrees by more than two standard errors. Most likely, the transverse de-focusing is implemented correctly in simulations but other uncertainties have not been properly resolved, as mentioned in [2].

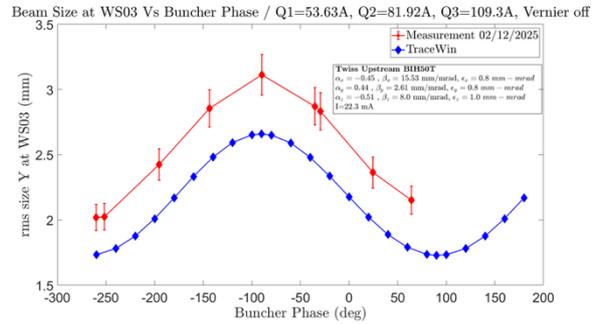

Figure 9: Comparing bunch transverse size in data and simulation with extracted Twiss parameters.

## CONCLUSION

Several longitudinal measurements were carried out to quantify beam in the Transition section. Buncher and Vernier synchronous phase and cavity amplitude were extracted. Bunch length was measured at varying Buncher phase and longitudinal emittance and Twiss parameters were calculated in TraceWin simulation.